\documentstyle[12pt]{article}

\input{tcilatex}
\begin{document}

\begin{titlepage}
\begin{center}
{\hbox to\hsize{
\hfill \bf HIP-2001-43/TH }}
{\hbox to\hsize{\hfill August 2001 }}

\bigskip
\vspace{3\baselineskip}

{\Large \bf

Composite quarks and leptons in higher space-time dimensions\\}

\bigskip

\bigskip

{\bf M. Chaichian$^{\mathrm{a}}$, 
J.L. Chkareuli$^{\mathrm{b}}$ and A. Kobakhidze$^{\mathrm{a,b}}$ \\}
\smallskip

{ \small \it
$^{\mathrm{a}}$High Energy Physics Division, Department of Physics, University of Helsinki and \\
Helsinki Institute of Physics, FIN-00014 Helsinki, Finland\\
$^{\mathrm{b}}$Andronikashvili Institute of Physics, Georgian Academy of Sciences, 380077 Tbilisi, Georgia\\}

\bigskip

\vspace*{.5cm}

{\bf Abstract}\\
\end{center}
\noindent
A new approach towards the composite structure of quarks and leptons
in the context of  the higher dimensional unified theories is proposed. Owing to the certain strong  dynamics, 
much like an ordinary QCD, 
every  possible  vectorlike set of composites appears
in higher dimensional bulk space-time, however,  through a proper Sherk-Schwarz compactification
only chiral
multiplets of composite quarks and leptons survive as the massless states
in four dimensions. In this scenario restrictions related with the 't Hooft's anomaly matching condition
are turned out to be  avoided and, as a result, the composite models look rather simple and
 economic.
We demonstrate our approach by an explicit  construction of model of preons and their
composites unified in the supersymmetric $SU(5)$ GUT
in five space-time dimensions. The model predicts exactly three families of the composite 
quarks and leptons being the triplets of the chiral horizontal symmetry $SU(3)_h$ which 
automatically appears  in the composite spectrum when going to ordinary four dimensions.
\bigskip

\bigskip

\end{titlepage}
\baselineskip=16pt

\section{Introduction}

The observed replication of quark-lepton families and hierarchy of their
masses and mixings are one of the major puzzles of modern particle physics.
In this respect it is conceivable to think that quark and lepton
spectroscopy finds its ultimate explanation in terms of the subfermions
(preons) and their interactions in analogy with an explanation of hadronic
spectroscopy in the framework of the quark model. However, a direct
realization of this program seems to meet serious difficulties. Among the
problems appeared the basic one is, of course, that related with the
dynamics responsible for a production of the composite quarks and leptons
whose masses $m_{f}$ are turned out to be in fact much less than a
compositeness scale $\Lambda _{C}$ which must be located at least in a few
TeV region to conform with observations \cite{data}. Indeed, if, as usual,
one considers underline preon theory to be QCD-like, then one inevitably
comes to the vectorlike bound state spectra where most naturally $m_{f}\sim
\Lambda _{C}$. To overcome this difficulty one has to require the presence
of some chiral symmetry which being respected by the strong preon dynamics
makes quark and lepton bound states to be massless. As 't Hooft first argued 
\cite{1}, such a chiral symmetry to be preserved in the spectrum of massless
composite fermions must yield the same chiral anomalies as those appearing
in the underline preon theory. However, this anomaly matching condition is
turned out to be too restrictive to drive at the physically interesting
self-consistent models. As a result, most of existing models \cite{2,3,4}
are rather complex and controversial and often contain too many exotic
composite states apart from the ordinary quarks and leptons.

Supersymmetric preon models \cite{3,4} follow to somewhat different pattern
of the anomaly matching condition since in this case the physical
composites, quarks and leptons, appear as both the three-fermion
(''baryons'') and scalar-fermion (''mesons'') bound states. More
interestingly, these models may provide a new dynamical alternative for
obtaining light composite fermions, which emerge as (quasi)Goldstone
fermions \cite{3} when the starting global symmetry $G$ of the
superpotential is spontaneously broken down to some lesser symmetry $H$. In
the recent years, there was renewal of interest in supersymmetric preon
models \cite{4} based on a powerful technique developed within the strongly
interacting $N=1$ supersymmetric gauge theories \cite{5}. However, despite
these attractive features of supersymmetric theories the supersymmetric
composite models generally suffer from the same drawbacks as the more
traditional non-supersymmetric ones.

In this Letter we suggest a new approach towards the composite structure of
quarks and leptons proposing a presence of extra space-time dimensions at
the small distances comparable or a bit larger than a radius of
compositeness $R_{C}\sim 1/\Lambda _{C}$. It is well known that the
compactification of extra space-time dimensions (depending on the detail of
dimensional reduction) was happened quite successful to get realistic four
dimensional models where supersymmetry \cite{6,7,8}, gauge symmetry \cite{9,
9e} and certain discrete symmetries such as $P$ and $CP$ \cite{10} are
broken in an intrinsically geometric way. Following to this line of
arguments we find that owing to a certain Scherk-Schwarz compactification 
\cite{6} the composite quarks and leptons are turned out to be massless in
four dimensions, while all unwanted states (residing in the bulk) are
massive. In this way the restrictions related with an original 't Hooft
anomaly matching condition can be avoided. Thereby, the physical composite
models look rather simple and economic as we will show shortly by a few
examples of the elementary preons and their composite unified in the $SU(5)$
SUSY GUT initially appearing in five space-time dimensions (5D).

\section{Supersymmetry in 5D and Scherk-Schwarz compactification}

Before turning to the construction of composite models let us recall some
aspects of the $N=1$ 5D supersymmetry and Scherk-Schwarz compactification
which are relevant for our subsequent discussion. Consider in 5D the $N=1$
supersymmetric gauge theory with a local symmetry group $G$ under which the
matter fields transform according to one of its irreducible representation $%
R $. The $N=1$ supersymmetric Yang-Mills supermultiplet ${\cal V}=(A^{M}$, $%
\lambda ^{i}$, $\Sigma $, $X^{a})$ in 5D contains a vector field $%
A^{M}=A^{M\alpha }T^{\alpha }$, a real scalar field $\Sigma =\Sigma ^{\alpha
}T^{\alpha }$, two gauginos $\lambda ^{i}=\lambda ^{i\alpha }T^{\alpha }$,
which form a doublet under the $R$-symmetry group $SU(2)_{R}$, and auxiliary
fields $X^{a}=X^{a\alpha }T^{\alpha }$ being a triplet of $SU(2)_{R}$ (here $%
M=0,1,2,3,4$ are space-time indices; $i=1,2$ and $a=1,2,3$ are $SU(2)_{R}$%
-doublet and $SU(2)_{R}$-triplet indices, respectively; $\alpha $ runs over
the $G$ group index values and $T^{\alpha }$ are generators of $G$ algebra).
These fields are combined into the $N=1$ 4D vector supermultiplet $V=(A^{m}$%
, $\lambda ^{1}$, $X^{3})$ ($m=0,1,2,3$) and a chiral supermultiplet $\Phi
=(\Sigma +iA^{4}$, $\lambda ^{2}$, $X^{1}+iX^{2})$. The matter fields are
collected in the hypermultiplet ${\cal H}=(h^{i}$, $\Psi $, $F^{i})$ which
contains the scalar fields $h^{i}$ being a doublet of $SU(2)_{R}$, Dirac
fermion $\Psi =(\psi _{1}$, $\psi _{2}^{+})^{T}$ being the $SU(2)_{R}$
singlet, and also the $SU(2)_{R}$ doublet of auxiliary fields $F^{i}$. All
those fields form two $N=1$ 4D chiral multiplets, $H=(h^{1}$, $\psi _{1}$, $%
F^{1})$ and $H^{c}=(h^{2}$, $\psi _{2}$, $F^{2})$ transforming according to
the representations $R$ and anti-$R$ of gauge group $G$, respectively. The
5D supersymmetric and $G$-symmetric action then can be then written as (
see, e.g. \cite{11}): 
\begin{eqnarray}
S &=&\int d^{5}x\int d^{4}\theta \left[ H^{c}e^{V}H^{c+}+H^{+}e^{V}H\right] +
\nonumber \\
&&\int d^{5}x\int d^{2}\theta \left[ H^{c}\left( \partial _{4}-\frac{1}{%
\sqrt{2}}\Phi \right) H+h.c.\right] .  \label{2.1}
\end{eqnarray}
The above theory (\ref{2.1}) is in fact vectorlike and, hence, anomaly-free.

Now let us compactify the extra fifth dimension $x^{4}$ on a circle of
radius $R_{C}$. Aside from the trivial (periodic) boundary conditions under
the $2\pi R_{C}$ translation of extra dimension one can impose to the 5D
fields the following non-trivial ($U(1)$-twisted) ones: 
\begin{eqnarray}
H(x^{m},x^{4}+2\pi R_{C},\theta ) &=&\exp (i2\pi q_{H})H(x^{m},x^{4},e^{i\pi
(q_{H}+q_{H^{c}})}\theta ),  \nonumber \\
H^{c}(x^{m},x^{4}+2\pi R_{C},\theta ) &=&\exp (i2\pi
q_{H^{c}})H(x^{m},x^{4},e^{i\pi (q_{H}+q_{H^{c}})}\theta ),  \nonumber \\
V(x^{m},x^{4}+2\pi R_{C},\theta ,\overline{\theta })
&=&V(x^{m},x^{4},e^{i\pi (q_{H}+q_{H^{c}})}\theta ,e^{-i\pi
(q_{H}+q_{H^{c}})}\overline{\theta }),  \nonumber \\
\Phi (x^{m},x^{4}+2\pi R_{C},\theta ) &=&\Phi (x^{m},x^{4},e^{i\pi
(q_{H}+q_{H^{c}})}\theta ),  \label{2.2}
\end{eqnarray}
where $q_{H}$ and $q_{H^{c}}$ are the $R$ charges of the superfields $H$ and 
$H^{c}$, respectively. Due to the periodicity conditions (\ref{2.2}) the
component fields are Fourier expanded as: 
\begin{eqnarray}
h^{1}(x^{m},x^{4}) &=&\sum_{n=-\infty }^{\infty
}e^{ix^{4}(n+q_{H})/R_{C}}h^{1(n)}(x^{m}),  \nonumber \\
\text{ }h^{2}(x^{m},x^{4}) &=&\sum_{n=-\infty }^{\infty
}e^{ix^{4}(n+q_{H^{c}})/R_{C}}h^{2(n)}(x^{m})  \nonumber \\
\psi _{1}(x^{m},x^{4}) &=&\sum_{n=-\infty }^{\infty }e^{ix^{4}\left( n+\frac{%
q_{H}-q_{H^{c}}}{2}\right) /R_{C}}\psi _{1}^{(n)}(x^{m}),\text{ }  \nonumber
\\
\psi _{2}(x^{m},x^{4}) &=&\sum_{n=-\infty }^{\infty }e^{ix^{4}\left( n-\frac{%
q_{H}-q_{H^{c}}}{2}\right) /R_{C}}\psi _{2}^{(n)}(x^{m})  \nonumber \\
\lambda ^{1}(x^{m},x^{4}) &=&\sum_{n=-\infty }^{\infty }e^{ix^{4}\left( n-%
\frac{q_{H}+q_{H^{c}}}{2}\right) /R_{C}}\lambda ^{1(n)}(x^{m}),\text{ } 
\nonumber \\
\lambda ^{2}(x^{m},x^{4}) &=&\sum_{n=-\infty }^{\infty }e^{ix^{4}\left( n+%
\frac{q_{H}+q_{H^{c}}}{2}\right) /R_{C}}\lambda ^{2(n)}(x^{m})  \nonumber \\
A^{m}(x^{m},x^{4}) &=&\sum_{n=-\infty }^{\infty }e^{ix^{4}\frac{n}{R_{C}}%
}A^{m(n)}(x^{m}),\text{ }  \nonumber \\
\left( \Sigma +iA^{4}\right) (x^{m},x^{4}) &=&\sum_{n=-\infty }^{\infty
}e^{ix^{4}\frac{n}{R_{C}}}\left( \Sigma +iA^{4}\right) ^{(n)}(x^{m}).
\label{2.3}
\end{eqnarray}
Let us note now that all the fields with the non-trivial $R$-charges are
necessarily turned out to be massive when reducing the theory from 5D to 4D.
Particularly, zero modes of all the fermionic fields in (\ref{2.3}) and
those of the scalars $h^{1}$ and $h^{2}$ have a masses $\frac{q}{R_{C}}$,
where $q$ are the corresponding $R$ charges, while the zero modes of the
gauge fields $A^{m}$ and adjoint scalar $\left( \Sigma +iA^{4}\right) $ are
massless. Nevetheless, the latter picks up the mass of the order of $\sim 
\frac{1}{R_{C}}$ radiatively since the supersymmetry is broken by the above
Scherk-Schwarz compactification, so that in general only gauge fields $A^{m}$
are left to be massless. However, if the $R$ charges of the superfields $H$
and $H^{c}$ are equal ($q_{H}=q_{H^{c}}$) then, as one can quickly confirm
from (\ref{2.3}), the zero modes of $\psi _{1}$ and $\psi _{2}$ happen to be
massless as well. Note also that for the composite operators containing the
above superfields the $R$ charge assignment, and thus the spectrum of the
massless zero modes, can be rather different. This is a key point we will
use below in the construction of composite models of quarks and leptons.

\section{One-generation composite model}

Let us consider $N=1$ supersymmetric $G\otimes SU(N)_{HC}$ gauge theory in
5D, where $G$ is a gauged part of some hyperflavor symmetry $G_{HF}$, which
includes all the observed symmetries (color $SU(3)_{C}$ and electroweak $%
SU(2)_{W}\otimes U(1)_{Y}$, or grand unified symmetry $SU(5)$ etc), and $%
SU(N)_{HC}$, which describes hypercolor interactions responsible for the
formation of hypercolorless bound states from preons. We assume that the
preons and anti-preons are resided in the 5D hypermultiplets ${\cal P}%
=(P,P^{c})$ and transform under hypercolor gauge group $SU(N)_{HC}$ as its
fundamental ($P\sim N$) and anti-fundamental ($P^{c}\sim \overline{N}$)
representations, respectively. The preons should carry also the quantum
numbers related with the hyperflavor symmetry group $G_{HF}$ . The
hypercolor gauge group $SU(N)_{HC}$ has to be asymptotically free as is in
the case of an ordinary QCD. Otherwise, the theory will not be well defined
as an interacting quantum field theory (because of the Landau pole appeared)
and can be consistently treated only as a low energy limit of some other
theory. Thus, the asymptotic freedom of the $SU(N)_{HC}$ restricts a number
of the allowed hyperflavors $N_{HF}$ to be 
\begin{equation}
\frac{N_{HF}}{2}\leq N  \label{3.1}
\end{equation}

Now let us take $G$, the gauged part of a total hyperflavor symmetry $G_{HF}$%
, to be the minimal grand unified group, i.e. $G\equiv SU(5)$, so that the
preons transform under the $SU(5)\otimes SU(N)_{HC}$ as: 
\begin{eqnarray}
P_{(5)} &\sim &(5,N),  \nonumber \\
P_{(\overline{5})}^{c} &\sim &(\overline{5},\overline{N}),  \nonumber \\
P_{(s)i} &\sim &(1,N),  \nonumber \\
P_{(\overline{s})}^{ci} &\sim &(1,\overline{N}),  \label{3.2}
\end{eqnarray}
where $i=1,...,N_{g}$. Therefore, the total number of flavors is $%
N_{HF}=5+N_{g}$. The $SU(5)$ singlet preons (anti-preons) $P_{(s)i}(P_{(%
\overline{s})}^{ci})$ in (\ref{3.2}) are actually necessary in order to
produce the entire set of composite quark and leptons transforming as $%
\overline{5}+10$ representations of $SU(5)$. We call them ``generation''
preons. Thus, the preons carry all ``basic'' quantum numbers presently
observed in quark-lepton phenomenology at low energies, such as three
colors, two weak isospin componens (being unified within the $SU(5)$) and
the generation numbers as well.

Within the framework described above the minimal possible hypercolor group
is $SU(3)_{HC}$ which admits a single ($N_{g}=1$) ``generation'' preon and,
thus, in total only six hyperflavors of preons, $N_{HF}=6$. This hypercolor
interaction is assumed to be responsible for the formation of hypercolorless
``baryons'' 
\begin{eqnarray}
\overline{D}_{1} &\sim &P_{(5)}P_{(5)}P_{(5)}\sim \overline{10},\hspace{0.5cm%
}D_{1}\sim P_{(\overline{5})}^{c}P_{(\overline{5})}^{c}P_{(\overline{5}%
)}^{c}\sim 10,  \nonumber \\
D_{2} &\sim &P_{(5)}P_{(5)}P_{(s)}\sim 10,\hspace{0.5cm}\overline{D}_{2}\sim
P_{(\overline{5})}^{c}P_{(\overline{5})}^{c}P_{(\overline{s})}^{c}\sim 
\overline{10}  \label{3.3}
\end{eqnarray}
and ``mesons'' 
\begin{eqnarray}
\overline{Q} &\sim &P_{(\overline{5})}^{c}P_{(s)}\sim \overline{5},\hspace{%
1.4cm}Q\sim P_{(5)}P_{(\overline{s})}^{c}\sim 5,  \nonumber \\
M &\sim &P_{(\overline{5})}^{c}P_{(5)}\sim 24+1,\hspace{0.5cm}S\sim P_{(%
\overline{s})}^{c}P_{(s)}\sim 1  \label{3.4}
\end{eqnarray}
at a compositeness scale $\Lambda _{C}$ (antisymmetrized products in (\ref
{3.3}) are meant). All these bound states come out in vectorlike $SU(5)$
representations and they are in fact the $N=1$ 4D superfields.

As in the previous section, compactifying the extra dimensions on a circle
of radius $R_{C}$ (and assuming that $R_{C}>1/\Lambda _{C}$ ) we impose
Scherk-Schwarz boundary conditions to the preonic superfields (\ref{3.2}) of
type 
\begin{eqnarray}
P_{(5)}(x^{m},x^{4}+2\pi R_{C},\theta ) &=&e^{i2\pi
q_{5}}P_{(5)}(x^{m},x^{4},e^{i\pi (q_{5}+q_{\overline{5}})}\theta ), 
\nonumber \\
P_{(5)}(x^{m},x^{4}+2\pi R_{C},\theta ) &=&e^{i2\pi
q_{5}}P_{(5)}(x^{m},x^{4},e^{i\pi (q_{5}+q_{\overline{5}})}\theta ), 
\nonumber \\
P_{(s)}(x^{m},x^{4}+2\pi R_{C},\theta ) &=&e^{i2\pi
q_{s}}P_{(s)}(x^{m},x^{4},e^{i\pi (q_{s}+q_{\overline{s}})}\theta ), 
\nonumber \\
P_{(\overline{s})}^{c}(x^{m},x^{4}+2\pi R_{C},\theta ) &=&e^{i2\pi q_{%
\overline{s}}}P_{(\overline{s})}^{c}(x^{m},x^{4},e^{i\pi (q_{s}+q_{\overline{%
s}})}\theta ),  \label{3.5}
\end{eqnarray}
where 
\begin{equation}
q_{5}+q_{\overline{5}}=q_{s}+q_{\overline{s}}.  \label{3.6}
\end{equation}
The vector supermultiplets and the adjoint superfields are periodic as in (%
\ref{2.2}). Expanding the 5D preonic fields as in (\ref{2.3}) one can see
that all fermionic preons are massive in 4D, thus low energy preonic theory
can be treated as a consistent quantum theory since the gauge anomalies are
absent. Obviously, supersymmetry is broken by the above boundary conditions (%
\ref{3.5}). Specifying the boundary conditions for the preonic fields one
can easily obtain $R$-charges for the composite states (\ref{3.3}) and (\ref
{3.4}) as well: 
\begin{eqnarray}
\overline{D}_{1} &\sim &3q_{5},\text{ }D_{1}\sim 3q_{\overline{5}},\text{ }%
D_{2}\sim 2q_{5}+q_{s},\text{ }\overline{D}_{2}\sim 2q_{\overline{5}}+q_{%
\overline{s}},  \nonumber \\
\overline{Q} &\sim &q_{\overline{5}}+q_{s},\text{ }Q\sim q_{5}+q_{\overline{s%
}},\text{ }M\sim q_{\overline{5}}+q_{5},\text{ }S\sim q_{\overline{s}}+q_{s}.
\label{3.7}
\end{eqnarray}
Since the $R$-charges (\ref{3.7}) for the composite states differ from those
of preons (\ref{3.5}), one can expect different spectrum of composite zero
modes. Particularly, we are looking for such an assignment of preonic $R$%
-charges (\ref{3.5}) which lead to massless composite fermions in 4D in $(%
\overline{5}+10)$ representation of $SU(5)$ that are nothing but composite
quarks and leptons. It is evident from (\ref{3.3}) and (\ref{3.4}) that we
should identify the fermionic components of $\overline{Q}$ superfield with
an anti-quintet of $SU(5)$ where down-type anti-quark and lepton doublet are
resided. The $SU(5)$ decuplet where quark doublet, up-type antiquark and
charged anti-lepton are resided can be identified with fermionic components
of either $D_{1}$ or $D_{2}$ superfields. The fermionic zero modes of $%
\overline{Q}$ and $D_{1}$ will be massless if the $R$-charges (\ref{3.5})
along with the equation (\ref{3.6}) satisfy also the following equations: 
\begin{eqnarray}
q_{\overline{5}}+q_{s} &=&\frac{q_{\overline{5}}+q_{5}}{2}  \label{3.8} \\
3q_{\overline{5}} &=&\frac{q_{\overline{5}}+q_{5}}{2}.  \label{3.9}
\end{eqnarray}
Solving the equations (\ref{3.6}, \ref{3.8}, \ref{3.9}) one has to remember
that due to periodicity $R$-charges $q$ are defined up to an arbitrary
integer number, $q=q+k,$ $k\in {\bf Z}$. To ensure that only a desired set
of fermionic zero modes are massless in 4D we restrict general $U(1)$%
-twisted boundary conditions to some discrete $Z_{K}$ ones. It is easy to
verify then that any $K\neq 2,3,4,6,9,12$ will provide the desired solutions
of (\ref{3.6}, \ref{3.8}, \ref{3.9}): 
\begin{equation}
q_{5}=\frac{5}{K},\text{ }q_{\overline{5}}=\frac{1}{K},\text{ }q_{s}=\frac{2%
}{K},\text{ and }q_{\overline{s}}=\frac{4}{K}.  \label{3.10}
\end{equation}
The minimal choice is $Z_{5}$-twisted boundary conditions with $R$-charges $%
q_{5}=0,$ $q_{\overline{5}}=\frac{1}{5},$ $q_{s}=\frac{2}{5},$ and $q_{%
\overline{s}}=-\frac{1}{5}$, so that only one generation of composite quarks
and leptons are massless in 4D at low energies. All extra composite states
are massive with masses of the order of the order of $\frac{1}{R_{C}}$.

If one identifies the quark-lepton decuplet of $SU(5)$ with fermionic
components of $D_{2}$ superfield then one has to determine the $R$-charges
from the equations (\ref{3.6}, \ref{3.8}) and the equation 
\begin{equation}
2q_{5}+q_{s}=\frac{q_{\overline{5}}+q_{5}}{2}\text{ },  \label{3.11}
\end{equation}
instead of (\ref{3.9}), appears. Any $Z_{K}$-twisted boundary conditions
with $K\neq 2,3,4,6,9,12$ and 
\begin{equation}
q_{5}=-\frac{2}{K},\text{ }q_{\overline{5}}=-\frac{4}{K},\text{ }q_{s}=\frac{%
1}{K},\text{ and }q_{\overline{s}}=-\frac{7}{K}  \label{3.12}
\end{equation}
will lead to the desired solutions. The minimal possibility is again $Z_{5}$
but now with the following $R$-charges: $q_{5}=-\frac{2}{5},$ $q_{\overline{5%
}}=\frac{1}{5},$ $q_{s}=\frac{1}{5},$ and $q_{\overline{s}}=-\frac{2}{5}$.
It looks quite intriguing that just composite quarks and leptons (without
any extra states) unified within the $SU(5)$ gauge theory emerge at low
energies in 4D from a simple and economic preon dynamics discussed above.

\section{Three-generation composite model}

One can easily extend the above model with one generation of composite
quarks and leptons to the case of three composite generations by simply
copying the above structure thrice, thus resulting in a model with
hypercolor group $SU(N)_{1}\otimes SU(N)_{2}\otimes SU(N)_{3}$. However, a
more interesting way is based on treating the $SU(5)$-singlet preons in (\ref
{3.2}) as the carriers of quantum numbers associated with quark-lepton
generations. Thus we will take three ``generation'' preons (anti-preons) $%
P_{(s)i}$ ($P_{(\overline{s})}^{ci}$) ($i=1,2,3$) and the global $%
SU(3)_{P_{(s)}}$ symmetry of 5D preonic Lagrangian will be interpreted as a
``horizontal'' hyperflavor symmetry $SU(3)_{h}$ for quark-lepton families
(see below). Therefore, we will also require this symmetry to be survived
upon the Scherk-Schwarz compactification, that is to say, the $R$-charges
for all three ``generation'' preons are the same. Now altogether there are $%
N_{HF}=8$ hyperflavors of preons and thus, due to the asymptotic freedom
constraint (\ref{3.1}), the minimal hypercolor group is $SU(4)_{HC}$. While,
following to arguments used in the previous section such a composite model
leading to three quark-lepton generations can easily be constructed, it
seems to be more interesting to take the $SU(5)_{HC}$ as the hypercolor
group. Apart from the possibility to treat all the massles composites in the
same way as the pure baryonic composites, this case may be of a special
interest as the case suggesting some starting extra hypercolor-hyperflavour
symmetry ($HC\leftrightarrow HF$) in the 5D. The composite ``baryons'' and
``mesons'' are then:

\begin{eqnarray}
\overline{D}_{1} &\sim &P_{(5)}P_{(5)}P_{(5)}P_{(s)}P_{(s)}\sim (\overline{10%
}{\bf ,}\overline{3})\text{ },\hspace{0.8cm}D_{1}\sim P_{(\overline{5}%
)}^{c}P_{(\overline{5})}^{c}P_{(\overline{5})}^{c}P_{(\overline{s})}^{c}P_{(%
\overline{s})}^{c}\sim (10,3),  \nonumber \\
D_{2} &\sim &P_{(5)}P_{(5)}P_{(s)}P_{(s)}P_{(s)}\sim (10,1)\text{ },\hspace{%
0.8cm}\overline{D}_{2}\sim P_{(\overline{5})}^{c}P_{(\overline{5})}^{c}P_{(%
\overline{s})}^{c}P_{(\overline{s})}^{c}P_{(\overline{s})}^{c}\sim (%
\overline{10},1),  \nonumber \\
\overline{Q} &\sim &P_{(5)}P_{(5)}P_{(5)}P_{(5)}P_{(s)}\sim (\overline{5},3)%
\text{ },\hspace{0.9cm}Q\sim P_{(\overline{5})}^{c}P_{(\overline{5})}^{c}P_{(%
\overline{5})}^{c}P_{(\overline{5})}^{c}P_{(\overline{s})}^{c}\sim (5,%
\overline{3}{\bf ),}  \nonumber \\
S &\sim &P_{(5)}P_{(5)}P_{(5)}P_{(5)}P_{(5)}\sim (1,1)\text{ },\hspace{0.9cm}%
\overline{S}\sim P_{(\overline{5})}^{c}P_{(\overline{5})}^{c}P_{(\overline{5}%
)}^{c}P_{(\overline{5})}^{c}P_{(\overline{5})}^{c}\sim (1,1{\bf )}
\label{4.1}
\end{eqnarray}
and 
\begin{eqnarray}
\overline{Q}^{^{\prime }} &\sim &P_{(\overline{5})}^{c}P_{(s)}\sim (%
\overline{5},3)\text{ },\hspace{1.7cm}Q^{^{\prime }}\sim P_{(5)}P_{(%
\overline{s})}^{c}\sim (5,\overline{3})  \nonumber \\
M &\sim &P_{(\overline{5})}^{c}P_{(5)}\sim (24+1,1)\text{ },\hspace{0.9cm}%
I\sim P_{(\overline{s})}^{c}P_{(s)}\sim (1,8+1{\bf )},  \label{4.2}
\end{eqnarray}
respectively, transforming under $SU(5)\otimes SU(3)_{h}$ as indicated in
brackets (anti-symmetrization of all the $SU(5)$ and $SU(3)_{h}$ indices are
meant in (\ref{4.1})). One can see that the $SU(5)$ decuplets
(anti-decuplets) in (\ref{4.1}), being the triplets (anti-triplets) and
singlets of the global family symmetry $SU(3)_{h}$, are pure baryonic
composites. As to the $SU(5)$ anti-quintets (quintets), being triplets
(anti-triplet) of the $SU(3)_{h}$, they appear as both baryonic (\ref{4.1})
and mesonic (\ref{4.2}) composites. Also some other states, singlets and
adjoints of $SU(5)$ and $SU(3)_{h}$, appear in the composite spectrum (\ref
{4.1},\ref{4.2}). Now, as soon as the fermionic zero modes proposed for the $%
D_{1}$ supermultiplet in (\ref{4.1}) are massless, one has to ensure that
the zero modes of fermionic components of the baryonic anti-quintet $%
\overline{Q}$ in (\ref{4.1}) or mesonic anti-quintet $\overline{Q}^{^{\prime
}}$ in (\ref{4.2}) (but not of the both) are also massless in order the low
energy composite model to be anomaly-free, thus giving an unique assignments
of the massless composites to the representation $(\overline{{\bf 5}}+10,3)$
of the $SU(5)\otimes SU(3)_{h}$. Remarkably, one can come to this basic
consequence even if starts with an arbitrary number of the generation preons 
$P_{(s)i}$ ($P_{(\overline{s})}^{ci}$) ($i=1,2,...,N_{g}$). Since, according
to the above construction (\ref{4.1},\ref{4.2}), a number of the composite $%
SU(5)$ decuplets is given by the $N_{g}(N_{g}-1)/2$, while the composite
anti-quintets by the number $N_{g}$ by itself (whether they are the baryonic
or meson composites), one is unavoidably come to the $SU(5)$ anomaly
cancellation condition of type

\begin{equation}
\frac{N_{g}(N_{g}-1)}{2}=N_{g}  \label{e1}
\end{equation}
from which immediately follows that $N_{g}=3$. Thus the above model actually
predicts three full generations of composite quarks and leptons being the
triplets of the chiral global family symmetry $SU(3)_{h}$ automatically
appeared in the composite spectrum.

Proceeding as in the previous section, one can easily determine the desired $%
R$-charges. If we identify the composite quarks and leptons with fermionic
zero modes of the baryonic composites $D_{1}$ and $\overline{Q}$ in (\ref
{4.1}) then preonic $R$-charges along with the equation (\ref{3.6}) must
satisfy the following equations:

\begin{eqnarray}
3q_{\overline{5}}+2q_{\overline{s}} &=&\frac{q_{\overline{5}}+q_{5}}{2}
\label{4.3} \\
4q_{5}+q_{s} &=&\frac{q_{\overline{5}}+q_{5}}{2}  \label{4.3e}
\end{eqnarray}
The desired solutions is provided by $Z_{6}$-twisted boundary conditions
(which is the minimal one) with $R$-charges defined as: 
\begin{equation}
q_{5}=q_{\overline{5}}=\frac{1}{6},\text{ }q_{\overline{s}}=-\frac{1}{6}%
\text{ and }q_{s}=\frac{1}{2}.  \label{4.4}
\end{equation}

In the case when the composite $SU(5)$ decuplets are identified with
fermionic zero modes of the baryonic composite $D_{1}$ (\ref{4.1}), while
the composite anti-quintet with the mesonic composite $\overline{Q}%
^{^{\prime }}$(\ref{4.2}) one should replace the equation (\ref{4.3e}) by
the equation (\ref{3.8}). It is easy to verify that the minimal solution
will be once again provided by $Z_{6}$-twisted boundary conditions but now
with the following $R$-charges: 
\begin{equation}
q_{5}=q_{\overline{5}}=\frac{1}{6},\text{ }q_{\overline{s}}=\frac{1}{3}\text{
and }q_{s}=0.  \label{4.6}
\end{equation}
Remarkably, only three generations of composite quarks and leptons emerge as
a massless states, while all other composites are massive, thus decoupling
from the low-energy particle spectrum.

\section{Discussion and conclusion}

Some questions concerning the dynamics of the composite models discussed
above must be further elaborated. The major ones are: How the $SU(5)$ and
subsequently the electroweak symmetries are broken? How the masses for
composite quarks and leptons are generated? Can one naturally explain the
hierarchies of masses and mixings of composite quarks and leptons? Here we
will briefly outline some possible scenarios one can think about.

In fact, one can use the $SU(5)$-adjoint superfield $\Phi $ to break $SU(5)$
symmetry down to the $SU(3)_{C}\otimes SU(2)_{W}\otimes U(1)_{Y}$ Standard
Model gauge group . In the supersymmetric uncompactified limit there are
degenerate flat vacuum directions for the scalar component of $\Phi $. Among
these vacua one can certainly find the $SU(5)$--breaking and $%
SU(3)_{C}\otimes SU(2)_{W}\otimes U(1)_{Y}$--invariant one. In such a vacuum
the preons will acquire $SU(5)$ non-invariant masses but this does not
affect their subsequent dynamics resulting in formation of composite states.
The degeneracy of vacuum states of course are lifted when one takes into
account supersymmetry breaking effects due to the Scherk-Schwarz
compactification. Alternatively, one can break $SU(5)$-symmetry through the
condensation of the scalar components of composite mesonic superfield $M$ (%
\ref{3.4},\ref{4.2}). Similarly, to break $SU(2)_{W}\otimes U(1)_{Y}$
electroweak symmetry one can use the doublet (anti-doublet) components of
composite quintets (anti-quintets). Since, the supersymmetry is broken, one
inevitably faces with gauge hierarchy problem which can be resolved by
fine-tuning as in the usual non-supersymmetric GUTs. Alternatively, one can
think that the solution to the gauge hierarchy problem appears due to the
strong renormalization of the electroweak Higgs mass which is driven to an
infrared stable fixed-point of the order of electroweak scale, while being
of the order of GUT scale at higher energies \cite{12}. Relatively large
extra dimensions play crucial role in this scenario by inducing fast
(power-law) evolution of gauge and Yukawa couplings.

The same mechanism could explain the observed hierarchies of quark-lepton
masses and mixings along the lines discussed in \cite{13}. These scenarios
can be actually operative in the case of composite quarks and leptons as
well. However, following to a more traditional way, one can think that the
hierarchy of quark-lepton masses and mixings are related with spontaneous
breaking of the global chiral $SU(3)_{h}$ horizontal symmetry appeared in
our model together the three quark-lepton generations predicted. That is, as
one can presently think, the main benefit of the above consideration.
Actually, the chiral horizontal symmetry $SU(3)_{h}$ is known \cite{14} to
work successfully both in quark and lepton sector and can readily be
extended to the composite quarks and leptons as well. It would be
interesting to gauge this symmetry within the preon model. However, a direct
gauging of the chiral horizontal symmetry typically leads to the $SU(3)_{h}$
triangle anomalies in the effective 4D theory. One way to overcome this
problem is to introduce some extra massless states which properly cancel
these anomalies in a traditional way. Another, and perhaps more interesting
possibility, is to cancel 4D anomalies by Callan-Harvey anomaly inflow
mechanism \cite{15} assuming a presence of 4D hypersurface (3-brane) in 5D
bulk space-time where the composite quarks and leptons are localized.

From purely phenomenological point of view it is certainly interesting to
study whether the compositeness scale, as well as the compactification one,
can be lowered down to the energies accessible for the high energy
colliders. Of course, these and related issues deserve more careful
investigation.

Various extensions of the simple models presented here are also interesting
to study. One can consider different gauge groups and more extra dimensions
as well. Particularly, one can study the possibility to unify the $SU(5)$
symmetry with the gauged horizontal $SU(3)_{h}$ and/or hypercolor $%
SU(N)_{HC} $ symmetries within a single gauge group (for earlier attempts
see, e.g. \cite{16}). It is certainly interesting to investigate the
dynamical emergence of gauge symmetries themselves with the composite gauge
bosons within the approach undertaken in this paper. And finally, from more
fundamental point of view it could be encouraging to study string theories
where the string excitations are identified with preons rather than the
physical quarks and leptons (for earlier discussion, see \cite{pati}).

To conclude, we have proposed a new approach towards the quark and lepton
compositeness within the higher dimensional unified theories where owing to
proper Scherk-Schwarz compactification the composite quarks and leptons are
turned out to be massless in four dimensions, while all unwanted states
(residing in the bulk) are massive. The prototype models discussed here are
rather simple and economic, so we think this approach will help to construct
the largely realistic composite models of quarks and leptons in a not
distant future.

\section*{Acknowledgements}

This work was supported by the Academy of Finland under the Project 163394.
One of us (JLC) would like to acknowledge a warm hospitality during his
visit to High Energy Physics Divison, Department of Physics, University of
Helsinki where part of this work was done.

\end{document}